\documentclass[%
reprint,
showpacs,preprintnumbers,
amsmath,amssymb,
aps,
longbibliography,
]{revtex4-1}

\usepackage{graphicx}% Include figure files
\usepackage{dcolumn}% Align table columns on decimal point
\usepackage{bm}% bold math
\usepackage{epstopdf}
\usepackage{xcolor}
\usepackage[utf8]{inputenc}
\usepackage{sistyle}
\usepackage[T1]{fontenc}
\SIthousandsep{,}
\usepackage{baskervald}
\usepackage[T1]{fontenc}
\usepackage{rotating}
\usepackage[margin=1in]{geometry}
\usepackage{gensymb}
\usepackage{amsmath}
\usepackage{float}
\begin{document}
	
	\preprint{APS/123-QED}

\title{The Plasma Membrane is Compartmentalized by a Self-Similar Cortical Actin Meshwork}% Force line breaks with \\

\author{Sanaz Sadegh$^1$}
\author{Jenny L. Higgins$^2$}
\author{Patrick C. Mannion$^2$}
\author{Michael M. Tamkun$^{3,4}$}
\author{Diego Krapf$^{1,2}$}

\affiliation{ $^1$Department of Electrical and Computer Engineering, Colorado State University, Fort Collins, Colorado 80523, USA \\ $^2$ School of Biomedical Engineering, Colorado State University, Fort Collins, Colorado 80523, USA\\ $^3$ Department of Biomedical Sciences, Colorado State University, Fort Collins, Colorado 80523, USA\\ $^4$ Department of Biochemistry and Molecular Biology, Colorado State University, Fort Collins, Colorado
80523, USA.}%Lines break automatically or can be forced with \\

\date{\today}% It is always \today, today,

\begin{abstract}
	
A broad range of membrane proteins display anomalous diffusion on the cell surface. Different methods provide evidence for obstructed subdiffusion and  diffusion on a fractal space, but the underlying structure inducing anomalous diffusion has never been visualized due to experimental challenges. We addressed this problem by imaging the cortical actin at high resolution while simultaneously tracking individual membrane proteins in live mammalian cells. Our data confirm that actin introduces barriers leading to compartmentalization of the plasma membrane and that membrane proteins are transiently confined within actin fences. Furthermore, superresolution imaging shows that the cortical actin is organized into a self-similar meshwork. These results present a hierarchical nanoscale picture of the plasma membrane.

\end{abstract}

\pacs{87.15.K-, 87.15.Vv}

\maketitle % Insert title

\section{INTRODUCTION}	

The plasma membrane is a complex fluid where lipids and proteins continuously interact and generate signaling platforms in order to communicate with the outside world. One of the key mechanisms by which membrane molecules search reaction sites is based on lateral diffusion. Quantitative imaging methods, such as single-particle tracking \cite{murase2004ultrafine,kusumi2005paradigm,manley2008high,clausen2013visualization}, spatiotemporal image correlation spectroscopy \cite{di2013fast}, fluorescence correlation spectroscopy (FCS) \cite{lenne2006dynamic,ruprecht2011spot}, and STED-FCS \cite{honigmann2014scanning,andrade2015cortical}, show that the dynamics of proteins and lipids in the plasma membrane often deviate from normal diffusion. In particular, the mean square displacement (MSD) does not grow linearly in time as expected for Brownian motion \cite{barkai2012strange,hofling2013anomalous,metzler2014anomalous,krapf2015chapter}. This behavior suggests processes that hinder diffusion. Since the formation of protein complexes is governed by diffusion-mediated encounters, hindered diffusion plays fundamental roles in cell function.

Unveiling the underlying mechanisms leading to the observed anomalous diffusion on the cell membrane is critical to  understanding  cell behavior. Anomalous diffusion in the plasma membrane can be caused by macromolecular crowding \cite{dix2008crowding}, transient binding \cite{weigel2011PNAS}, heterogeneities \cite{ratto2003anomalous,manzo2015weak}, and membrane compartmentalization by the underlying cytoskeleton \cite{fujiwara2002phospholipids,kusumi2005paradigm,auth2009diffusion,andrade2015cortical}. In recent years it has become evident that a single mechanism cannot account for the complex dynamics observed in the plasma membrane \cite{krapf2015chapter}. We have shown that interactions with clathrin coated pits (CCPs) cause anomalous diffusion and ergodicity breaking \cite{weigel2011PNAS,weigel2013quantifying}. However, it was observed that this process coexisted with a different anomalous diffusion mechanism attributed to diffusion within a fractal topology. Experimental evidence for the organization of the plasma membrane by the cortical actin cytoskeleton has been provided by measurements in cell blebs, spherical protrusions that lack actin cytoskeleton \cite{tank1982enhanced}, and in the presence of actin-disrupting agents \cite{lavi2012lifetime,gudheti2013actin,andrade2015cortical}. The picket-fence model explains these observations by postulating that the mobility of membrane-bound molecules is hindered by the actin-based cytoskeleton in close proximity to the plasma membrane, leading to transient confinement \cite{sheetz1983membrane,saxton1997single,kusumi2005paradigm}. 
Confinement and segregation of membrane components can have important physiological consequences by allowing the formation of functional domains on the cell surface. 
However, in spite of the vast evidence that has accumulated over the last two decades, a direct observation of the dynamic compartmentalization of membrane proteins by underlying actin fences is challenging due to the spatial and temporal resolutions required for its visualization.

Here we employ superresolution imaging and single-particle tracking of membrane proteins to elucidate the compartmentalization of the plasma membrane by intracellular structures. While tracking individual voltage-gated potassium channels as described in our previous studies \cite{weigel2011PNAS}, we found that these membrane proteins exhibited anomalous diffusion on the cell surface. We now report that the anticorrelated dynamics are best modeled by obstructed diffusion instead of fractional Brownian motion and we directly visualize the transient confinement of potassium channels by cortical actin in live cells. In order to characterize the cortical actin meshwork, we employ stochastic optical reconstruction microscopy (STORM) to obtain superresolution images in fixed cells. We find a non-integer fractal dimension for the actin cortex and a broad distribution of compartment sizes as expected for a self-similar structure. These observations consistently explain the anticorrelated subdiffusive motion of membrane proteins and provide new insights on the hierarchical organization of the plasma membrane.

\section{Results}

\subsection{Kv1.4 and Kv2.1 ion channels undergo subdiffusion in the plasma membrane} 

\begin{figure*}
 	{\includegraphics[width=16.4 cm]{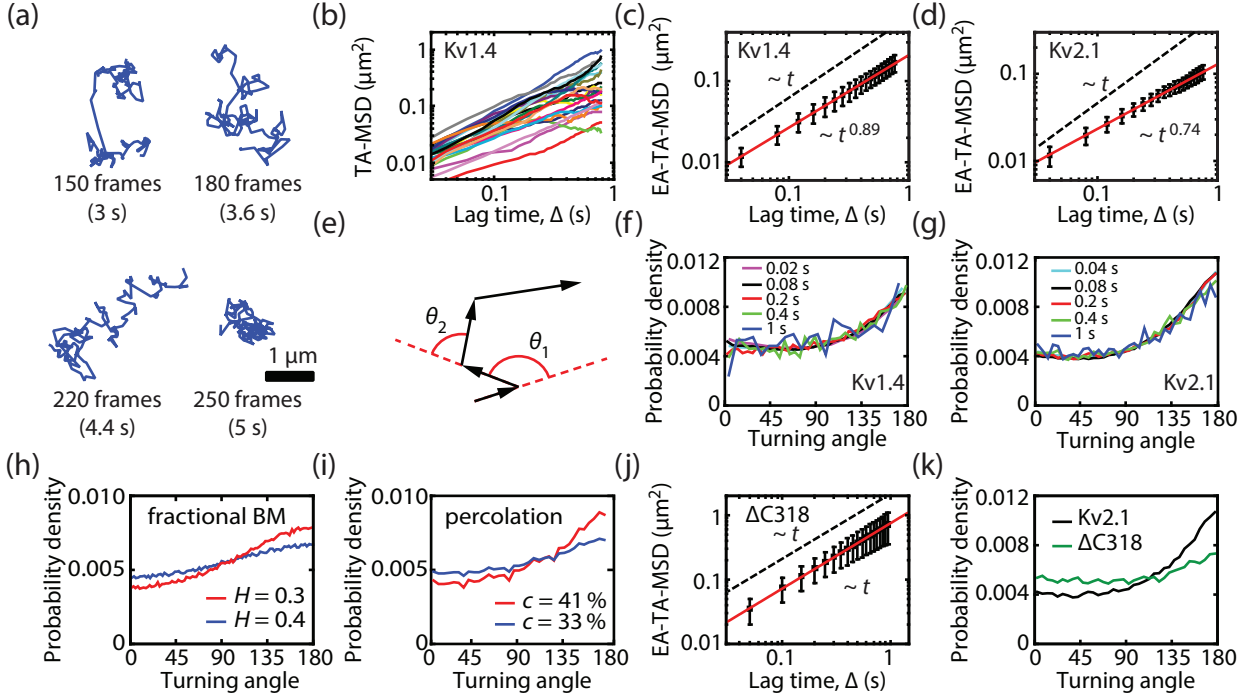}}
 	\caption{Voltage-gated potassium channels Kv1.4 and Kv2.1 undergo subdiffusion in the plasma membrane. (a) Four Kv1.4 representative trajectories obtained by single-particle tracking. (b)  Time-averaged MSD (TA-MSD) as a function of lag time $\Delta$ for 20 individual Kv1.4 trajectories. (c) Ensemble-averaged time-averaged MSD (EA-TA-MSD) averaged over 1,312 Kv1.4 trajectories ($n=10$ cells). (d)  EA-TA-MSD averaged over 6,385 Kv2.1 trajectories ($n=14$ cells). The dashed lines in  c and d are visual guides for linear behavior (free diffusion), i.e., $\left\langle\overline{\delta^2(\Delta)}\right\rangle \sim \Delta$. Error bars show standard deviation. (e) Sketch illustrating the construction of turning angles from a particle trajectory.  (f)-(g) Turning angle distributions for  Kv1.4 (10 cells, 1,312 trajectories) and Kv2.1 (14 cells, 6,385 trajectories). Turning angle distributions are constructed for lag times between 20 ms and 1 s. (h) Turning angle distributions for fractional Brownian motion simulations with Hurst exponents 0.3 and 0.4. (i) Turning angle distribution for simulations of obstructed diffusion with obstacle concentrations 33\% and 41\%, i.e., site percolation. (j) MSD averaged over 3,114 $\Delta$C318 trajectories ($n=5$ cells). (k) Turning angle distributions for Kv2.1 and $\Delta$C318 (5 cells, 3,114 trajectories) measured with lag time of 200~ms.}
 	\label{fig:turn}
 \end{figure*}

Voltage-gated potassium channels Kv1.4 and Kv2.1 were expressed in human embryonic kidney (HEK) cells, labeled with quantum dots (QDs) \cite{weigel2011PNAS}, and imaged using total internal reflection fluorescence (TIRF) microscopy at 50 frames/s, so that individual molecules could be detected on the cell surface. 
Kv1.4 and 2.1 are similar in size, 654 and 853 amino acids, respectively, but share less than 20\% overall amino acid identity \cite{deal1996molecular}. They are placed into distinct gene subfamilies because of this low identity. They are most similar within a central core domain composed of six transmembrane alpha helices and the ion conducting pore. In contrast, they share no amino sequence identity within the cytoplasmic N- and C-terminal regions; each Kv1.4 subunit has 402 cytoplasmic amino acids while the Kv2.1 subunits have 624. Both channels exist as homotetrameric structures giving the functional channel 24 membrane spanning domains and a total of either 1608 or 2496 cytoplasmic amino acids.
Figure~\ref{fig:turn}(a) shows representative  trajectories of Kv1.4 channels. The motion of the ion channels was initially evaluated in terms of their time-averaged MSD,
\begin{equation}
\overline{\delta^2(\Delta)}=\frac{1}{T-\Delta}\int_{0}^{T-\Delta} |\textbf{r}(t+\Delta)-\textbf{r}(t)|^2 \textrm{d}t,
\end{equation}
where $T$ is the total experimental time, $\textbf{r}$ the particle position, and $\Delta$ the lag time, i.e., the time difference over which the MSD is computed. When a particle displays Brownian diffusion, the MSD is linear in lag time, i.e., $\overline{\delta^2(\Delta)} \sim \Delta$. In contrast, anomalous diffusion is characterized by a different MSD scaling, namely  MSD $\sim \Delta^\alpha$, where $\alpha$ is the anomalous exponent. Anomalous diffusion is classified as subdiffusion when $0<\alpha < 1$ and superdiffusion when $\alpha > 1$. Figure~\ref{fig:turn}(b) shows the MSD of 20 individual trajectories. The MSDs of Kv1.4 as well as Kv2.1 channels show subdiffusive behavior, albeit with large apparent fluctuations.  Figures~\ref{fig:turn}(c) and~\ref{fig:turn}(d) show the  MSDs averaged over 1,312 Kv1.4 ($n=10$ cells) and 6,385 Kv2.1 ($n=14$ cells) trajectories, respectively, $\left\langle\overline{\delta^2(\Delta)}\right\rangle$. Throughout the manuscript we employ overlines to denote time averages and brackets to denote ensemble averages. The anomalous exponent $\alpha$ of Kv1.4 was found to be 0.89 and that of Kv2.1 was 0.74, indicating subdiffusion in both cases.
  
Several distinct mathematical models lead to subdiffusion \cite{hofling2013anomalous,metzler2014anomalous,krapf2015chapter}. Among the most well-accepted types of subdiffusion in biological systems, we encounter (i) obstructed diffusion, (ii) fractional Brownian motion (fBM), and (iii) continuous time random walks (CTRW). 
Both fBM \cite{mandelbrot1968fractional, deng2009ergodic} and obstructed diffusion \cite{saxton1994anomalous, ben2000diffusion, weigel2012obstructed} are models for subdiffusive random walks with anticorrelated increments that have been extensively used in live cells. fBM describes the motion in a viscoelastic fluid \cite{szymanski2009elucidating,ernst2012fractional}, which can be caused by macromolecular crowding \cite{guigas2007degree,weiss2013single}. fBM is a generalization of Brownian motion that incorporates correlations with power-law memory. It is characterized by a Hurst exponent $H$ that translates into an anomalous exponent $\alpha=2H$. Obstructed diffusion describes the motion of a particle hindered by immobile (or slowly moving) obstacles, e.g., percolation. As the concentration of immobile obstacles increases, the availability of space decreases. Near a critical concentration known as percolation threshold, the obstacles form a fractal with dead ends in all length scales. In particular, the reduction of the available space results in anomalous diffusion with a recurrent exploration pattern. A CTRW is a generalization of a random walk where a particle waits for a random time between steps \cite{scher1975anomalous}. When the waiting times are asymptotically distributed according to a power law such that the mean waiting time diverges, the CTRW is subdiffusive. 
These three models describe very distinct physical underlying mechanisms but they can yield similar sublinear MSD scaling, particularly in obstructed diffusion and fBM models. Thus the MSD analysis is insufficient to elucidate the type of random walk. 

Different tests beyond the MSD have been employed to distinguish among types of subdiffusive random walks, including \textit{p}-variations \cite{magdziarz2009fractional}, first passage probability distribution \cite{rangarajan2000first}, mean maximal excursion \cite{tejedor2010quantitative}, Gaussianity \cite{weiss2004anomalous}, and fractal dimensions \cite{meroz2013test}. Here we employ the distribution of directional changes, i.e., the turning angles, a tool that probes correlations in the particle displacements and has been shown to contain information on the complexity of a random walk \cite{burov2013distribution}. Figure~\ref{fig:turn}(e) illustrates the construction of turning angles from a particle trajectory. In simple Brownian motion, the turning angles are uniformly distributed. Contrastingly, when the steps are correlated the distribution of turning angles is not uniform \cite{burov2013distribution}. Figures~\ref{fig:turn}(f) and~\ref{fig:turn}(g) show the distribution of turning angles of Kv1.4  and Kv2.1 for different lag times (1,312 Kv1.4 tracks, 10 cells and 6,385 Kv2.1 tracks, 14 cells). Both distributions peak at $\theta = 180^{\circ}$ indicating the particles are more likely to turn back than to move forward. In other words, Kv channels have a preference to go in the direction from where they came rather than to persist moving in the same direction. This property is a fingerprint of subdiffusive random walks with anticorrelated increments. Besides the shape of the distribution, the dependence on lag time bears valuable information. Strikingly, we observe that the distribution is independent of lag time, i.e., we measure the same distribution of directional changes whether the lag time is 20 ms or 1 s.

We examined numerical simulations of fBM and obstructed diffusion and found that they have distinctive attributes in their distribution of directional changes. Figure~\ref{fig:turn}(h) shows the distribution of directional changes for subdiffusive fBM simulations with Hurst exponents $H=0.3$ and $0.4$. Even though the distributions peak at $180^{\circ}$, the probability density function is different from the experimental data [Figs.~\ref{fig:turn}(f) and~\ref{fig:turn}(g)]. In our experimental data, the turning angle distributions increase sharply as $\theta$ approaches $180^{\circ}$ and most of the deviations from a uniform distribution are above $90^{\circ}$. However, fBM gives rise to a gradual increase that takes place mainly in the range $45^{\circ} < \theta < 135^{\circ}$. Further, the turning angles of fBM reach a plateau, in contrast to our measurements. Conversely, obstructed diffusion, strongly resembles our experimental results. Figure~\ref{fig:turn}(i) shows the turning angle distribution for obstructed diffusion simulations in a square lattice with obstacle concentrations $33\%$ and $41\%$ \cite{weigel2012obstructed}. Note that $41\%$ is slightly above the percolation threshold. These results show that the motion of Kv channels in the plasma membrane is better modeled by percolation, i.e., obstructed diffusion, rather than motion in a viscoelastic medium, i.e., fBM. 
    
Potential obstacle candidates for obstructed diffusion in the plasma membrane are the cortical cytoskeleton, lipid rafts, and  extracellular glycans. By evaluating the MSD and turning angle distribution of $\Delta$C318, a mutant in which the last 318 amino acids of the C-terminus of Kv2.1 channel had been deleted \cite{vandongen1990alteration}, we found that the anticorrelated diffusion originates from interactions with intracellular structures. We observed that $\Delta$C318 channels diffuse freely in the plasma membrane, $\alpha=1$ with a diffusion coefficient $D=0.19\ \mu \textrm{m}^2$/s [Fig.~\ref{fig:turn}(j), n=3114 tracks, 5 cells]. Further, the distribution of turning angles of $\Delta$C318 was flattened, as expected for Brownian diffusion [Fig.~\ref{fig:turn}(k)], 
indicating the intracellular C terminal domain of Kv2.1 plays a key role in the anticorrelations within the particle trajectory. Even though the distribution of turning angles in the $\Delta$C318 mutant is close to that in Brownian motion, a small peak is still noticeable at $180^{\circ}$ suggesting additional complexities in the plasma membrane.

\begin{figure*}
	\centerline{\includegraphics[width=13.2 cm]{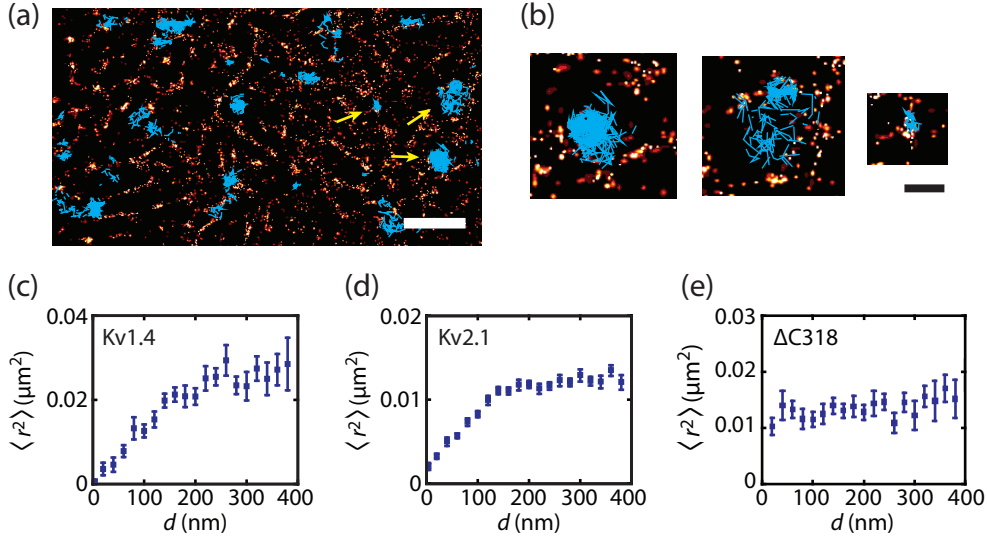}}
	\caption{Cortical actin transiently confines Kv channels. (a) Trajectories of individual Kv2.1 channels (shown in cyan) overlaid on actin PALM image (shown in red). Scale bar 2 $\mu$m. (b) Enlargements of the areas indicated with yellow arrows in a. Scale bar is 500 nm. The left trajectory shows confinement in a large compartment, the middle one shows hoping between two compartments and the right one shows confinement in a nanoscale domain. (c)-(e) Mean square displacements $\langle r^2 \rangle$ covered by Kv1.4 and Kv2.1 and $\Delta$C318 channels in 200 ms as a function of their maximum distance from nearest actin feature. Error bars indicate standard errors.}
	\label{fig:Effect}
\end{figure*}

In contrast to Kv1.4, which is homogeneously distributed on the cell membrane, a subpopulation of Kv2.1 channels forms micron-sized clusters 
that localize to endoplasmic reticulum (ER)-plasma membrane junctions \cite{fox2013plasma,fox2015induction}. 
Thus, we expect both the ER and the cortical cytoskeleton introduce intracellular interactions with Kv2.1 channels. 
To identify the origin of the observed anticorrelated diffusion, we analyzed the motion of non-clustered Kv2.1 channels, i.e., the channels that reside outside ER-plasma membrane junctions. We labeled Kv2.1 channels both with green fluorescent protein (GFP) and QDs \cite{fox2015induction}. While all the channels were labeled with GFP, only a small fraction included QDs in order to enable both single-particle tracking and cluster identification [supplementary Fig. S1]. We observed that the distribution of directional changes of non-clustered channels is indistinguishable from that of the overall population [supplementary Fig. S2]. Thus, we can exclude interactions with the ER as the 
cause for anticorrelated subdiffusion. 
These observations suggest that diffusion is hindered by intracellular components, possibly the cortical cytoskeleton, in agreement with a membrane-skeleton fence model \cite{kusumi2005paradigm}. 

We observed that the distribution of turning angles were independent of lag times [Figs.~\ref{fig:turn}(f) and~\ref{fig:turn}(g)] within the probed spatial and temporal scales. These observations indicated the anticorrelated subdiffusion of Kv channels did not have an evident characteristic time scale. This type of random walk is consistent with diffusion on a self-similar structure, i.e., a fractal subspace. In order to visualize the difference between diffusion on a fractal structure, and diffusion on a meshwork with a characteristic length scale, we performed simulations of motion of a particle in the presence of permeable fences that introduce compartments with a well-defined length scale [supplementary Fig. S3]. In these simulations, we observed that the distribution of turning angles is not time invariant; the peak at $180^{\circ}$ grows as we increase the lag-time up to a characteristic time, and then it decays when the lag-time increases further [supplementary Figs. S3(b) and S3(c)]. Thus, hop-diffusion with a narrow distribution of confinement sizes exhibits a time-dependent turning angle distribution (with a well-defined characteristic time scale), in contrast to our experimental results where the turning angle distribution is time-invariant.

\subsection{Cortical actin transiently confines Kv1.4 and Kv2.1 channels}

We observed that Kv channels undergo obstructed diffusion. The $\Delta$C318 mutant data indicated that hindering of the particle motion originated within cytoplasmic structures in close proximity to the plasma membrane, in agreement with previous experimental evidence of transient confinement by the actin-based cytoskeleton \cite{tank1982enhanced,tsuji1986restriction,edidin1994truncation,andrews2008actin,andrade2015cortical}. Thus we examined the cortical actin as a candidate for the observed obstructed diffusion in the plasma membrane.

We imaged the cortical actin in live HEK cells using the photoactivatable probe tdEosFP \cite{wiedenmann2004eosfp} via an actin binding peptide (ABP) that reversibly binds to F-actin \cite{izeddin2011super}. 
Previous studies showed that expression of ABP-tdEosFP does not affect the organization of the cytoskeleton \cite{izeddin2011super, riedl2008lifeact}. By activating a sparse subset of tdEosFP and individually localizing them with high precision, we generated photoactivated localization microscopy (PALM) images using localizations from 100 frames (2~s), yielding a smooth video of the dynamic actin meshwork (supplementary Video S1). Both excitation and photoactivation were implemented in total internal reflection fluorescence (TIRF) so that only the actin adjacent to the plasma membrane was imaged. The dissociation of ABP-tdEosFP occurs with a time constant on the order of 40~s \cite{izeddin2011super}, thus the exchange within 2-s imaging is negligible. 

Figure~\ref{fig:Effect}(a) shows a representative PALM reconstruction of actin. Although the number of localizations in 100 frames is not adequate to fully resolve the cortical actin and some faint fluorescent single-filament structures might be missed in the images, we could use the reconstructed PALM image to study the interactions of the potassium channels with the actin cortex in live cells. Previous breakthrough experiments have reported simultaneous imaging of cortical actin cytoskeleton and single-particle tracking \cite{andrews2008actin,treanor2010membrane,torreno2016actin}. Here, to the best of our knowledge, we perform  for the first time simultaneous single-particle tracking measurements and imaging cortical actin with superresolution.

In order to find out whether actin-delimited domains as identified by PALM hinder diffusion and compartmentalize the cell surface, we imaged and tracked Kv1.4 and Kv2.1 channels on the cell surface while simultaneously imaging the cortical actin. Channels often remained confined within the areas enclosed by actin indicating actin acted as a barrier to channel diffusion. Figures \ref{fig:Effect}(a) and \ref{fig:Effect}(b) show Kv2.1 channel tracks for one video overlaid on the last reconstructed image of the cortical actin. However, this visualization method suffers from overlaying long trajectories on a single reconstruction image of the actin meshwork. In addition to being constrained by actin structures, some trajectories exhibit confinement within small nanoscale domains that do not appear to be enclosed by actin. We have previously shown that Kv channels exhibit frequent immobilizations when the channels are captured within clathrin-coated pits \cite{weigel2013quantifying}. Thus, the cortical actin cytoskeleton is not the sole mechanism by which the mobility of Kv channels is hindered. In order to deal with these complexities, we evaluated the MSDs as a function of proximity to actin.     

Given that actin hinders channel motility, we expect the particles to explore smaller areas when they are confined within smaller compartments. To test the actin fence hypothesis, we overlaid channel trajectories on the corresponding PALM image of actin obtained in 2 s, a time scale in which the actin structure is fairly persistent [supplementary video 1], with a sliding time window 0.2 s. For example, trajectories of the channels from 0 to 2 s were overlaid on the first reconstructed actin frame and the trajectories in the interval 200 ms to 2.2 s were overlaid on the second actin PALM frame. Then we partitioned the trajectories into 200-ms intervals and classified each segment according to the maximum distance $d$ of the particle to the nearest actin feature, calculated using an Euclidean distance map algorithm. We evaluated the ensemble-averaged MSD $\langle r^2 \rangle$ of all the segments located at a specific distance away from actin, i.e., we averaged the squared displacements in 200 ms of the particles transiently located a given distance from actin. Figures~\ref{fig:Effect}(c) and~\ref{fig:Effect}(d) show the MSD as a function of distance-to-actin for Kv1.4 and Kv2.1. For both channels we observed that as molecules dwell closer to actin their MSD decreases.

As a control of our method, we performed the same analysis for  $\Delta$C318 channels. Because of the lack of the intracellular domain, these channels should not have any interaction with the cortical cytoskeleton. We observed that the MSD of $\Delta$C318 channels is independent of distance from actin [Fig.~\ref{fig:Effect}(e)], which demonstrates the effect of intracellular structure in the transient confinement of Kv channels.

\subsection{Characterization of cortical actin meshwork}

\begin{figure*}%[b]
		\centerline{\includegraphics[width=12.1 cm]{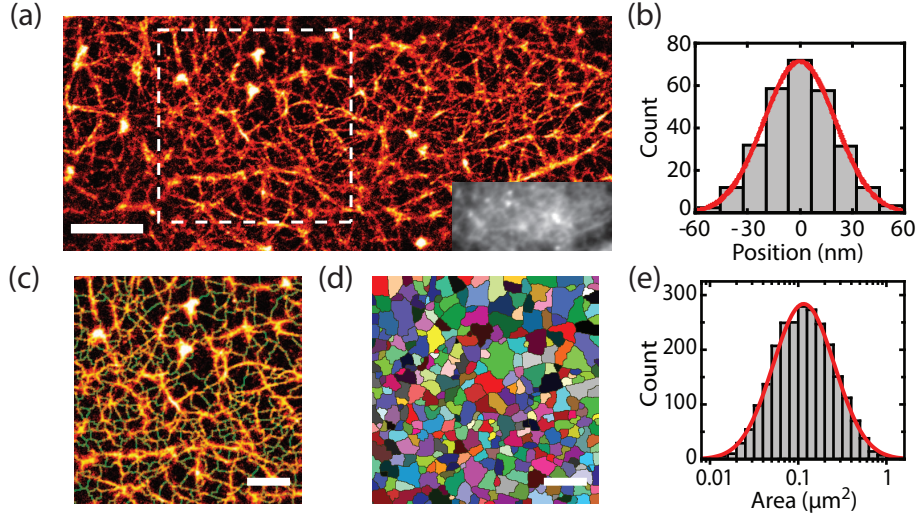}}
		\caption{ Characterizations of actin compartments. (a) Superresolution STORM image of the cortical actin in a HEK cell. The inset shows the conventional TIRF image. Scale bar is 2 $\micro$m. (b) Average cross-section profile of 20 filaments aligned by the center of each line. The red line is Gaussian fit with standard deviation $\sigma=20$~nm. (c) Watershed segmentation (shown in green) of the boxed area overlaid on the STORM image. (d) Compartments determined by watershed are designated with  different colors. Scale bars in c and d are 1 $\micro$m. (e) Distribution of compartment areas for fixed cells (9 cells, $n=$~2,500 compartments). Areas are shown in logarithmic scale and the red line is a log-normal distribution with shape parameter $\sigma=$0.8 $\micro$m.}
		\label{fig:STORM}
	\end{figure*}

When imaging live cells with PALM, the number of frames used in the reconstruction is restricted by the cell dynamic nature. A low number of frames results in insufficient detected particles to accurately determine the structure, also having a deleterious effect on resolution. In contrast, very high spatial resolution can be obtained in fixed cells by collecting data over long times \cite{rust2006sub,dempsey2011evaluation,xu2012dual}. Therefore, we used TIRF-STORM to visualize the compartments formed by cortical actin in fixed cells. Actin was labeled with phalloidin conjugated to Alexa Fluor 647, which binds actin filaments with high specificity without significantly enlarging them \cite{xu2012dual}. A total of 50,000 frames where used in the reconstruction. In our STORM reconstructions we observed both thick and thin actin structures, Fig~\ref{fig:STORM}(a). The finest structures that we observed had a cross section standard deviation of 20 nm (FWHM=48 nm). Figure~\ref{fig:STORM}(b) shows the average cross section profile of 20 lines aligned by the center of each line. The thickness of these lines in the reconstruction is governed by the localization accuracy, $20 \pm 8$ nm (mean $\pm$ SD, supplementary Fig. S4), which sets a lower bound on STORM resolution. Thus we are unable to determine whether these structures are individual filaments (10 nm in diameter) or actin bundles.

We employed a watershed segmentation algorithm \cite{meyer1990morphological} to identify actin-delimited compartments in the STORM reconstructions [Figs.~\ref{fig:STORM}(c) and \ref{fig:STORM}(d)] across the whole cell. The average percentage of the watershed meshwork covered by actin was $84 \pm 4 \%$ (mean $\pm$ SD, n=9 cells). Figure~\ref{fig:STORM}(e) shows the distribution of compartment areas ($n=2,500$ compartments). The areas of the compartments are fitted well by a log-normal distribution, which is a subexponential heavy-tailed distribution in the sense that it decays more slowly than any exponential tail \cite{kluppelberg1988subexponential}. The log-normal distribution is in good agreement with Kolmogorov's model for the distribution of particle sizes after repeated breakage \cite{kolmogorov1941logarithmically,epstein1947mathematical}. When a particle is divided into fragments in such a way that the fragment proportions are independent of the original particle size, a log-normal distribution emerges in the particle sizes after random repeated fragmentation. Analogously, actin-delimited compartments are split into smaller compartments by growing actin filaments and thus their distribution is predicted to be log-normal.

\begin{figure}%[b]
	\centerline{\includegraphics[width=8.4 cm]{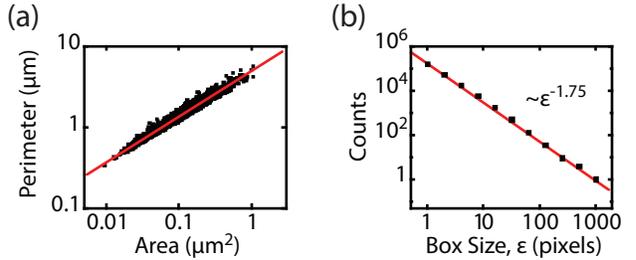}}
	\caption{Fractality of the cortical actin meshwork. (a) Log-log scatter plot of compartment perimeter vs. compartment area. The fitted line corresponds to $L=4.8 A^{0.55}$ (Pearson correlation coefficient $\rho=0.98$ in log scales). (b) Representative example of box counting algorithm in one cell where the exponent yields $d_f=1.75$.}
	\label{fig:Fractal}
\end{figure}

In addition to the compartment area distribution, the relation between perimeter and area contains valuable information. Perimeter-area relations have been extensively used to investigate the properties of complex planar shapes \cite{mandelbrot1983fractal,lovejoy1982area}. As expected, we observe that areas and perimeters of the different compartments are highly correlated [Fig.\ref{fig:Fractal}(a)]. This correlation indicates shape homogeneity among different compartments \cite{imre2004fractal}. Furthermore, the area exhibits the same scaling over the whole observed range, $A\sim L^{1.8}$, where $A$ and $L$ are compartment area and perimeter, indicating the same statistical character at different scales, and suggesting that compartments formed by cortical actin are scale-invariant in the observed range. Such scale invariance is a hallmark of a self-similar fractal structures.

Fractals are characterized by scaling properties governed by a non-integer dimension $d_f$, i.e., an anomalous dependence of the ``mass'' on the linear size of the system with $M\sim l^{d_f}$.
In a regular object such as a line, square, or cube, we would refer to its mass $M$ as the length, area or volume, respectively. In these regular cases the mass scales as $M\sim l^{d}$, where the $l$ is the linear size and $d=1,2,3$ is the spatial dimension. On the other hand, fractals such as a Sierpinsky gasket or a percolation cluster differ from Euclidean spaces and display a fractional dimension \cite{mandelbrot1983fractal,addison1997fractals}. 
Usually the capacity dimension is obtained using a box-counting algorithm that quantifies the mass scaling. In brief, the structure is placed on a grid, the number of occupied ``boxes'' are counted, and the process is iterated for finer grids. The number of occupied boxes scales as $N\sim\epsilon^{-d_f}$, where $\epsilon$ is the box length. Figure~\ref{fig:Fractal}(b) shows the computation of the fractal dimension of the cortical actin meshwork from the STORM image in a representative cell. The box counting analysis shows the actin structure exhibits statistical self-similarity over more than three decades.  Our data indicate the fractal dimension of the meshwork is $d_f=1.75\pm 0.02$ ($n=9$ cells).

\section{Discussion} 
Our current understanding of the plasma membrane is that of a complex partitioned fluid where molecules often undergo anomalous diffusion and can be segregated according to their function. We observe that K$^+$ channels perform a random walk with antipersistent nature, i.e., a random walk with an increased probability of returning to the site it just left. However elucidating the mechanisms that cause anomalous diffusion is not trivial because several different subdiffusion models lead to similar MSD scaling. The analysis of K$^+$ channel motion is further complicated by the occurrence of immobilizations with power law sojourn times, which introduce deviations from Gaussian functions in the distribution of displacements \cite{weigel2011PNAS,weigel2013quantifying}. Thus, we cannot employ Gaussianity-based tests to distinguish among complex antipersistent random walks. We find that the distribution of directional changes provides a robust test for the type of random walk. The measured channel trajectories are shown to be well described by obstructed diffusion but not by fBM. 

We observed that the Kv2.1 intracellular domain played a key role in the anomalous diffusion, in agreement with previous observations showing that 
the depth at which a membrane protein extends into the cytoplasm determined how frequently it encountered mechanical barriers \cite{edidin1994truncation}. The obvious candidate to obstruct the motion of proteins with large intracellular domains is the actin cytoskeleton. 
Thus, we visualized the cortical actin with high temporal and spatial resolution and evaluated its effect on membrane protein dynamics. Considering that some faint single-filament actin structures might not be accurately detected by PALM imaging, we can miss some interactions between actin and membrane proteins. Notwithstanding, we found that Kv channels are transiently confined by permeable actin fences, confirming existing models for membrane compartmentalization as an organizing principle of the actin cytoskeleton \cite{kusumi2005paradigm}.  By studying the diffusion of Kv2.1 channels outside ER-plasma membrane junctions, we verified that the observed subdiffusion is not due to interactions with ER. Other intracellular components such as intermediate filaments could also hinder protein diffusion and further compartmentalize the cell membrane, but these cytoskeletal filaments were not studied in the present work. Previous single-particle tracking works using lipids labeled with 40-nm gold nanoparticles have observed the compartmentalization of the plasma membrane of HEK293 cells, with 70-nm mean compartment size \cite{murase2004ultrafine}. However, this compartmentalization occurs with molecules having virtually no cytoplasmic domains and with a residence time close to 3 ms. At time scales above 50 ms, gold-labeled lipids were found to exhibit Brownian diffusion with an effective diffusion coefficient $D=0.41\  \mu\textrm{m}^2/\textrm{s}$ \cite{murase2004ultrafine}, similar to our observations for $\Delta$C318 Kv2.1 mutant.

Ion channels are observed to undergo anticorrelated anomalous diffusion over at least two orders of magnitude in time. In terms of percolation theory, this hints the cell surface is maintained close to criticality, i.e., near the percolation threshold. However, this hypothesis seems highly unlikely. 
A more feasible explanation stems from the emergence of a scale-invariant structure under the plasma membrane. 
We directly observed that, within the probed spatial scale, the actin cortex has in fact a self-similar nature. It is possible to speculate that actin fractality develops from its branching structure. 
Hierarchically branched structures have a fractal dimension $d_f$ such that $R_b=R_r^{d_f}$, where $R_b$ is the bifurcation ratio and $R_r$ is the length-order ratio \cite{newman1997fractal}. The bifurcation ratio can be interpreted as the average number of branches that emerge after a bifurcation and the length-order ratio is defined as the ratio between incoming branch length and the length of the emerging branches until the next bifurcation \cite{horton1945erosional}. Actin branching is driven by the Arp2/3 complex \cite{pantaloni2000arp2} with a bifurcation ratio $R_b=2$. Here we measure a meshwork fractal dimension $d_f=1.75$, which can arise from a branching pattern with $R_r=1.5$ or, in other words, the daughter branch being on average $1/3$ shorter than the mother branch. 
The fractal dimension of the cytoskeleton is in line with a broad range of fractal geometries found in biology ranging from the lung alveoli to subcellular structures such as mitochondrial membranes and the endoplasmic reticulum \cite{mandelbrot1983fractal}.
 
We propose that the fractal nature of the actin cortex is employed by the cell to organize the plasma membrane. The complexity of this structure leads to a hierarchical organization with domains in multiple length scales and the development of nested compartments. Such a dynamic hierarchical organization facilitates the active segregation of domains with different functions and the maintenance of reactants near reaction centers. Furthermore, the fractal nature of the cortical actin has broad implications for anomalous diffusion, for instance it could bridge the gap between the plasma membrane hop diffusion models and diffusion in a fractal that leads to anomalous dynamics over broad time scales. We foresee that a self-similar cytoskeleton structure also influences active actomyosin-mediated organization of the plasma membrane \cite{gowrishankar2012active} in such a way that these processes can take place over multiple length scales.

In conclusion, our findings show that the plasma membrane is compartmentalized in a hierarchical fashion by a dynamic cortical actin fractal. We find that the anomalous diffusion of  potassium channels is best modeled by obstructed diffusion or diffusion in a fractal. By combining PALM imaging with single-particle tracking we were able to directly visualize the hindering effect of cortical actin on the diffusion of the membrane proteins. We characterized the compartments formed by cortical actin using superresolution imaging in fixed cells and found evidence for the self-similar topology of this structure.

 \section{MATERIALS AND METHODS}
 
 \subsection{Cell transfection and labeling.} HEK 293 cells (passage 42-49; American Type Culture Collection) were cultured in phenol red Dulbecco’s Modified Eagle’s Medium (DMEM), supplemented with 10\% fetal bovine serum (FBS; Gibco) at 37 $^{\circ}$C. Cells were transfected to express a Kv2.1 or Kv1.4 construct with an extracellular biotin acceptor domain that, when coexpressed with a bacterial biotin ligase, results in biotinylated Kv channels on the cell surface \cite{o2006kv2}.  For live-cell actin imaging cells were transfected with 3 $\mu$g  of ABP-tdEos. ABP is the	actin-binding sequence of ABP140 from \textit{S. cerevisiae} consisting of 17 amino acids \cite{izeddin2011super, riedl2008lifeact}. Kv2.1-loopBAD-GFP (3 $\mu$g) was employed to identify Kv2.1 clusters on the cell surface. 
 
 For single-particle tracking, biotinylated channels were labeled with QDs. Cells were incubated for 10 minutes in HEK imaging saline with 1 nM streptavidin-conjugated QD705 or QD655 (Invitrogen) and 10 mg/mL bovine serum albumin (BSA) \cite{weigel2011PNAS} at 37 $^{\circ}$C. Following incubation the cells were rinsed again six times with HEK imaging saline to ensure the removal of any unbound QDs. The diameter of the QDs is in the range 10-15 nm, thus given that Kv channels are similar in size to the QDs, it is highly unlikely that the Kv:QD stoichiometry is higher than 1:1. We have previously shown that QD conjugation does not alter Kv channel diffusion \cite{weigel2011PNAS,weigel2013quantifying}. 
 
 \subsection{Live cell imaging.}Imaging was performed in an objective-based TIRF microscope built around an IX71 Olympus body. Sample temperature was kept at 37 $^{\circ}$C using objective and stage heater (Bioptechs). A 405-nm laser was used to activate tdEosFP fluorophore, while 473-nm and 561-nm lasers were used to excite it in its inactive and active states, respectively.

\subsection{Single-particle tracking.} QD labeling was controlled so that QDs remained at low density to allow for single-particle tracking \cite{weigel2011PNAS}. Particle detection and tracking were performed in MATLAB using u-track \cite{jaqaman2008robust}.
 
\subsection{Fixed cell imaging.} Cells were plated on Matrigel-coated 35 mm petri dishes. After 12 hours the cells were fixed and labeled with Alexa Fluor 647-phalloidin (Invitrogen). Image stacks were obtained using the same setup as live cell imaging. A continuous illumination of 638-nm laser was used to excite the Alexa Fluor 647. The laser power before the objective was 30 mW. To maintain an appropriate density of activated molecules, 405-nm laser was used in some experiments. 50,000 frames were collected to generate a superresolution image. 

\subsection{Fractal dimension.} The fractal dimension of the actin cortex was computed using a box-counting algorithm. The thresholded binary actin image was placed on a grid of square boxes of size $\epsilon$ and the number of occupied boxes was counted. The process was repeated for grids of boxes with different sizes. The number of occupied boxes scales as
\begin{equation}
 N\sim \epsilon^{-d_f},
 \end{equation}
where $d_f$ is the capacity dimension, or simply the fractal dimension.
 
\subsection{Statistics} Results show mean and s.e.m. unless indicated otherwise. All experimental results were obtained from multiple different dishes and days. The number of distinct imaging regions, i.e., different cells is indicated in the text.

\begin{acknowledgments}
We thank Keith Lidke and Sheng Liu for the codes for PALM reconstruction, Maxime Dahan for providing the plasmids to express ABP-tdEos, and Aubrey Weigel, Liz Akin, Mar\'{i}a Gracia Gervasi, Phil Fox, Ben Johnson, and Xinran Xu for useful discussions. We gratefully acknowledge the support of NVIDIA Corporation with the donation of a GeForce GTX TITAN used for this research. D.K. and J.L.H. thank Nikki Curthoys for technical advice with superresolution microscopy during the initial stages of the project. This work was supported by the National Science Foundation under Grant 1401432 (to DK) and the National Institutes of Health under Grant R01GM109888 (to MMT).
\end{acknowledgments}

%\bibliography{apssamp}

\providecommand{\noopsort}[1]{}\providecommand{\singleletter}[1]{#1}%

\end{document}